# Demonstration of KV-Class β-Ga$_2$O$_3$ Trench Junction Barrier Schottky Diodes with Space-Modulated Junction Termination Extension

Advait Gilankar[1,a)], Julian Gervassi-Saga[1], Martha R. McCartney[2], Nabasindhu Das[1], David Malcolm McComas[1], David J. Smith[2] and Nidhin Kurian Kalarickal[1]

[1]School of Electrical, Computer and Energy Engineering, Arizona State University, Tempe, AZ 85281
[2]Department of Physics, Arizona State University, Tempe, AZ 85287

[a)] Electronic mail: agilanka@asu.edu

In this work, we report on the design and fabrication of p-NiO/Ga$_2$O$_3$ trench junction barrier schottky diodes (JBSD) integrated with space-modulated junction termination extension (SM-JTE) and compare the performance with planar Ni/Ga$_2$O$_3$ schottky diodes (SBDs) and p-NiO/Ga$_2$O$_3$ heterojunction diodes (HJDs). The JBSDs achieved breakdown voltages exceeding 1.8 kV along with low leakage currents ($<10^{-2}$ A/cm$^2$), while displaying low turn on voltage ($V_{ON}$) of ~1V, which is similar to that of planar Ni/Ga$_2$O$_3$ SBDs. The fabricated devices showed excellent forward characteristics with low differential on-resistance ($R_{on,sp}$) ranging from 4-10.5 mΩ-cm$^2$, for fin width between 0.6-1.25 microns. Best performing device with fin width of 0.85μm showed a unipolar figure of merit (FOM) of ~0.7GW/cm$^2$. This work showcases the benefits of trench JBS design along with SM-JTE edge-termination for efficient high-performance kilovolt-class β-Ga$_2$O$_3$ diodes.

# I. INTRODUCTION

The growing demand for high performance, energy efficient power devices capable of operating at higher power densities and frequencies, as well as in harsh environments have compelled researchers to investigate semiconductor technologies beyond conventional materials such as Si, SiC and GaN.[1–3] In this context, ultra-wide bandgap semiconductors having a bandgap ($E_g$) larger than 3.4eV have attracted considerable attention over the past decade. Among UWBGs, $\beta$-Ga$_2$O$_3$ ($E_g$=4.8eV) has emerged as a highly promising candidate for applications in high-power electronics, owing to its high theoretical breakdown field (8 MV/cm) and wider range of n-type doping capability.[4–6] Additionally, the availability of large-area single-crystal substrates grown using bulk crystal growth techniques have enabled high quality homoepitaxy using various techniques such as Halide Vapor Phase Epitaxy (HVPE), Metal Organic Vapor Phase Epitaxy (MOVPE) and Molecular Beam Epitaxy (MBE).[7,8] High quality drift layers with low defect densities have enabled the demonstration high-performance small and large-area (>1mm$^2$) devices with several recent reports having surpassed the unipolar figure of merit (FOM) of SiC and GaN.[9,10]

Despite the above-mentioned advantages, $\beta$-Ga$_2$O$_3$ technology faces two key limitations: 1) the low-thermal conductivity of $\beta$-Ga$_2$O$_3$ and 2) the absence of reliable p-type doping to form p-n homojunctions.[4,5] While a $\beta$-Ga$_2$O$_3$ homo p-n junction may be impractical for power switching due to the large turn on voltage (~4.5 V), these junctions are essential for enabling advanced power device architectures such as junction barrier schottky diodes (JBSDs), D-MOSFETs and Trench-MOSFETs.[11,12] In addition, these junctions are also key to realizing high performance field terminations such as guard rings and junction termination, which are critical for achieving near theoretical breakdown performance.[13,14] Researchers have been successful in circumventing this drawback by using p-type oxide based hetero p-n junctions (such as p-NiO/n-Ga$_2$O$_3$ heterojunction).[15–18] In recent years, multiple groups have demonstrated p-NiO/ β-Ga$_2$O$_3$ heterojunction diodes (HJDs) utilizing edge terminations such as field plates, implant terminations, high-k dielectrics and JTEs (using p-NiO) achieving record high performance for 2-terminal devices.[15,19–21]

The primary drawback of p-NiO/ Ga$_2$O$_3$ HJDs is high (>2V) turn-on voltage which results in higher conduction losses in the on-state.[22,23] One way to overcome the challenge of high turn-on voltage is to fabricate junction barrier schottky diodes (JBSDs) which simultaneously offer benefits of low turn-on voltage and high reverse breakdown voltage. There have been prior reports employing the use of p-NiO on Ga$_2$O$_3$ to achieve high-performance JBSDs.[24–26] Additionally, trench Schottky barrier diodes (SBDs) have also been demonstrated in the past achieving high breakdown voltage, due to reduced surface field (ReSURF) effect at the Schottky metal/semiconductor junction.[10]

In this work, we demonstrate a deep trench-junction barrier schottky diode (JSBD) which simultaneously leverages the higher breakdown field of the NiO/Ga$_2$O$_3$ junction (at the trench corner) and the RESURF effect due to the trench design (at the Schottky contact) for enhanced breakdown voltage. In addition, the JBSDs are integrated with an effective space-modulated junction termination extension (SM-JTE) as the edge termination. The fabricated JBSDs display a low turn on voltage of ~1V, which is significantly lower than that of planar NiO/Ga$_2$O$_3$ HJDs fabricated on the same wafer. The diodes also achieve a high breakdown voltage between 1.8 and 2 kV, which is ~2× higher compared to planar NiO/Ga$_2$O$_3$ HJDs (without edge termination) and around ~4× higher compared to planar Ni/Ga$_2$O$_3$ SBDs (without edge termination).

## II.  EXPERIMENTAL AND DEVICE CHARACTERIZATION

The devices were fabricated on halide vapor phase epitaxy (HVPE) grown 10μm thick (001) β-Ga$_2$O$_3$ drift-layer procured from Novel Crystal Technologies (NCT), Japan, with doping concentration of ~1.1×10$^{16}$ cm$^{-3}$. The doping concentration in the substrate was estimated to be ~5×10$^{18}$ cm$^{-3}$. The sample underwent a standard solvent clean and HF dip for about 15 minutes, following which the alignment marks were patterned and etched into the sample using ICP-RIE dry-etch. This was followed by e-beam evaporation of Ti/Au (50/100nm) backside ohmic contacts and subsequent rapid thermal annealing (RTA) at 470˚C for 1 minute in N$_2$ environment. A tri-layer hard mask consisting of SiO$_2$ (300nm), SiN$_x$ (100nm) and Ni (100nm) was used to pattern Ga$_2$O$_3$ trenches. First the SiO$_2$ and SiN$_x$ layers under the Ni were etched away using a CHF$_3$/O$_2$ RIE etch. Afterwards, the Ga$_2$O$_3$ trenches were defined using BCl$_3$ gas-based ICP dry-etch process with ICP and RF power of 800W and 30W, respectively. The diodes were patterned to have different fin-widths (W$_f$) which varied from 0.6μm to 1.25 μm with an estimated trench depth of 0.85 μm. Post dry etching of trenches, the Ni layer in the tri-layer mask was removed using HCl. A controlled dip in diluted BOE (1:10) was used to create an undercut in the SiO$_2$/SiN$_x$ mask making use of the vastly different etch rates for these dielectrics in BOE. Following the dry-etch process, 150nm thick p-NiO was deposited using RF sputtering in 10% O$_2$/ Ar environment. NiO sputtering was carried out using an 8" NiO target at pressure of 4mT and RF power of 300W. Due to the undercut in the SiO$_2$/SiN$_x$ mask, the NiO deposited on the top-side of the trenches could be easily lifted off by performing a short dip in dilute BOE (1:10), leaving NiO only on the trench bottom and the sidewalls. The space-modulated JTE was formed by a two-step NiO deposition with 50nm and 100nm thickness. Following NiO sputtering, devices were RTP annealed in $N_2$ at 300°C for 300s. Finally, a thick metal stack Ni/Au (50nm/80nm) was evaporated to form the Schottky contact as well as the ohmic contact to the p-NiO. The process flow used for the fabrication of trench JBSDs and schematic showing critical steps is summarized in Fig. 1. Planar p-NiO/ $Ga_2O_3$ HJDs (thickness of NiO ~150nm) and Ni/$Ga_2O_3$ SBDs were also fabricated on the same epi-layer alongside the trench SM-JTE diodes. The cross-sectional schematic of the device after fabrication is shown in Fig. 2 (a)-(b). Post-fabrication, the devices were electrically characterized using Keysight B1500A parameter analyzer and B1505A high-voltage analyzer. Sample for TEM imaging was prepared by focused ion beam (FIB) milling using Thermo Fisher Helios 5 UX. The HR-TEM imaging of the cross-section of trench structure was performed using an image-corrected FEI Titan 80–300 operated at 300 kV. The cross-sectional TEM image of trench SM-JTE diode with fin-width of 0.6μm is shown in Fig. 2(c).

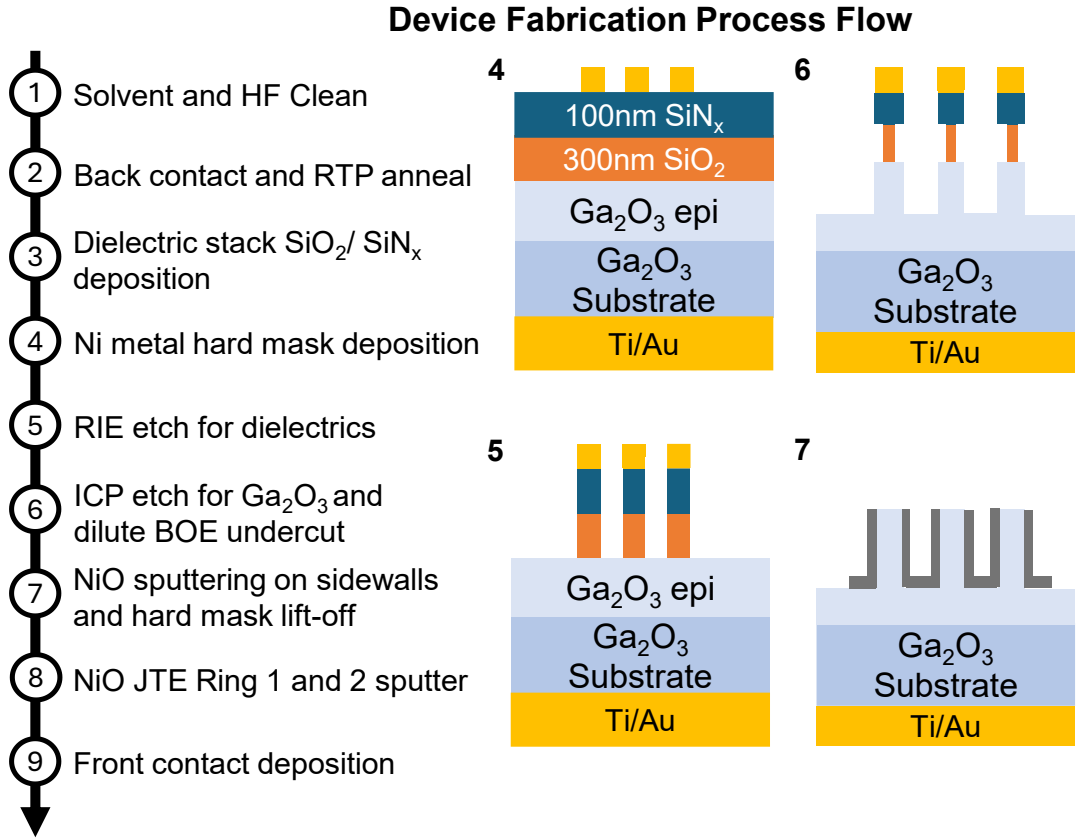

**Device Fabrication Process Flow**

1. Solvent and HF Clean
2. Back contact and RTP anneal
3. Dielectric stack SiO$_2$/ SiN$_x$ deposition
4. Ni metal hard mask deposition
5. RIE etch for dielectrics
6. ICP etch for Ga$_2$O$_3$ and dilute BOE undercut
7. NiO sputtering on sidewalls and hard mask lift-off
8. NiO JTE Ring 1 and 2 sputter
9. Front contact deposition

FIG. 1. Schematic of process flow used for device fabrication and schematic showing critical fabrication steps 4-7.

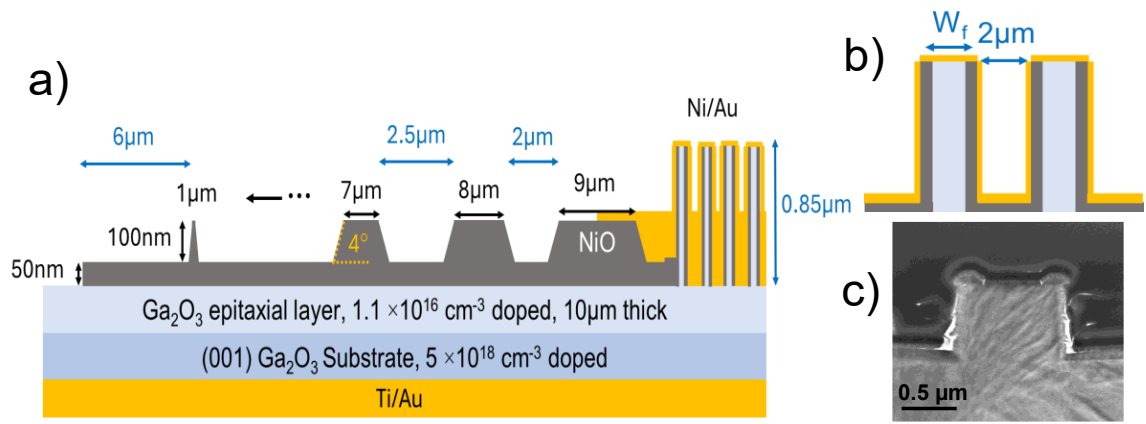

FIG. 2. Device schematic of deep trench SM-JTE JBSDs: **(a)** complete device schematic, **(b)** magnified cross-sectional schematic of the trench structure, and **(c)** Cross-sectional HR-TEM image of device with 0.6 μm fin-width.

# III. RESULTS AND DISCUSSION

The donor concentration in the $Ga_2O_3$ epilayer is estimated using C-V measurements as shown in Fig. 4(b). The net donor concentration ($N_D$–$N_A$) in the drift-layer is estimated to be $\sim 1.1 \times 10^{16}$ cm$^{-3}$. The forward current-voltage (I-V) characteristics of diodes with different fin-widths ($W_f$) are shown in Fig. 3 (a) and (b). The forward currents are normalized with respect to total fin area. The specific differential on-resistance ($R_{on,sp}$) for diodes with different fin-widths is shown in Fig. 3(c). The $R_{on,sp}$ (normalized to fin-widths) varies from about 4m$\Omega$-cm$^2$ ($W_f$=1.25$\mu$m) to 10.5m$\Omega$-cm$^2$ ($W_f$=0.6$\mu$m).

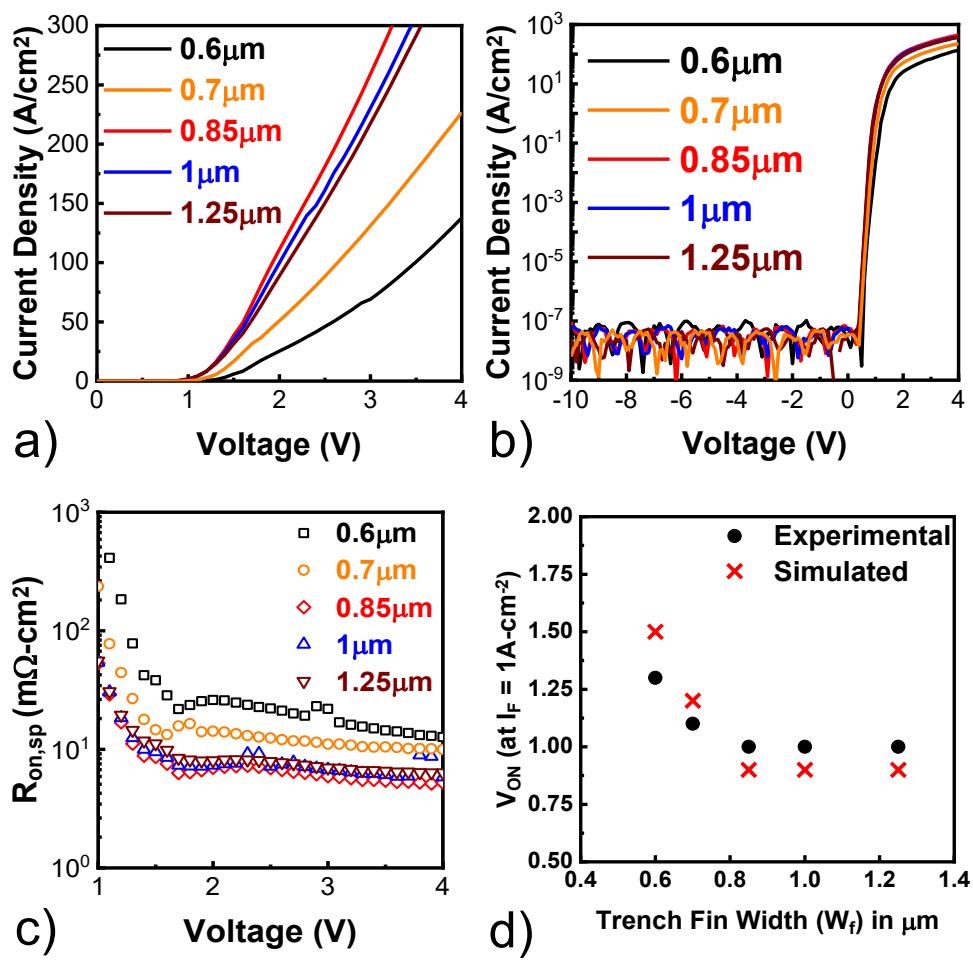

F$_{IG}$. 3. Forward I-V characteristic for trench diodes with different fin-widths in **(a)** linear scale, **(b)** semi-log scale. **(c)** Plot showing differential on-resistance ($R_{on,sp}$) versus voltage for different fin-widths (The forward I-V was normalized with respect to total fin-width area). **(d)** Comparison of turn-on voltage at current density of 1A-cm$^{-2}$ ($V_{ON}$) as a function of trench fin-width ($W_f$) extracted from simulated and experimentally measured forward I-Vs.

All the JBSDs display a low turn-on voltage between 1 and 1.3V (extracted at 1A-cm$^{-2}$), and the forward current density increases as the fin-width increases. In fig. 3(d), we observe that the turn-on voltage of JBSD decreases as the fin-width increases, with devices having $W_f > 0.85$ µm displaying a constant $V_{ON}$ of ~1 V. This trend is supported by TCAD simulations of the device structure under forward bias conditions. This suggest that devices with $W_f < 0.85$ µm do not offer optimal performance in terms of the turn on voltage. The forward turn-on characteristics of the trench JBS diodes were also compared with planar SBDs and HJDs as shown in Fig. 4 (a). The SBDs and trench JBSDs show nearly similar forward performance in terms of both turn-on voltage and forward current density. The SBD and trench JBSD ($W_f = 0.85$µm) show turn-on voltage of 0.8V and 1V respectively measured at a current density of 1 A-cm$^{-2}$. The NiO/Ga$_2$O$_3$ HJD, however, displays a much higher turn-on voltage of around 2.5V. In addition to higher turn-on voltage, the HJD is also more resistive and shows lower forward current density recorded at same voltage compared to trench JBSDs.

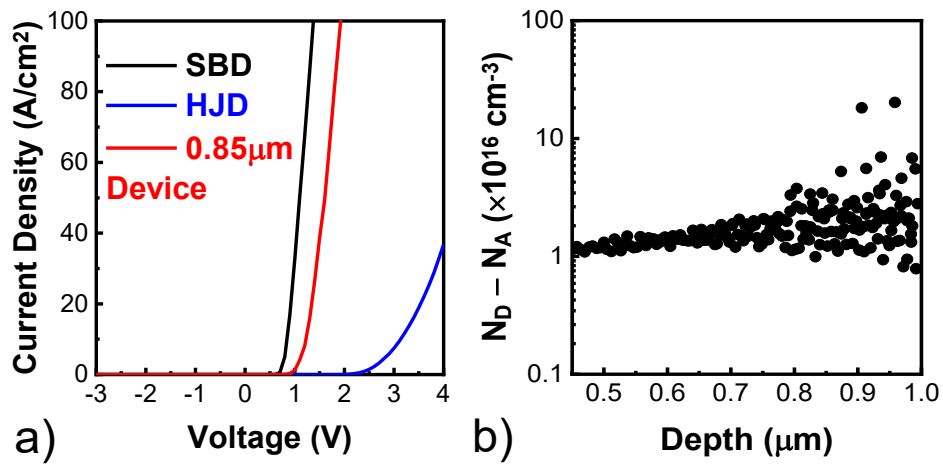

FIG. 4. **(a)** Comparison of forward I-V characteristics of Ni/Ga$_2$O$_3$ (SBD), (150nm thick NiO) NiO/Ga$_2$O$_3$ (HJD) and trench JBSD. **(b)** Net charge concentration extracted from C-V measurements is shown.

The measured reverse I-V characteristics of the devices (JBSDs, SBDs and HJDs) is shown in Fig. 5 (a). The reverse leakage current for the JBSDs is normalized to the total device area (active area +JTE area).

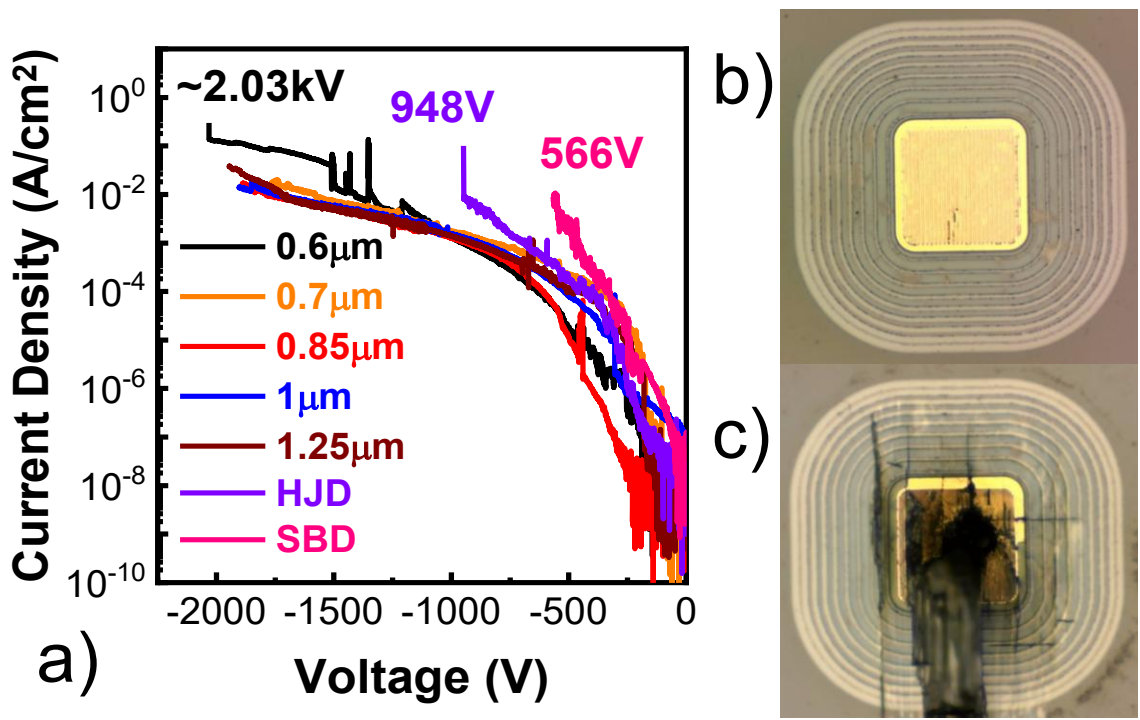

F<small>IG</small>. 5. **(a)** Reverse I-V characteristics for trench JBS diodes with different fin-widths, SBD, and HJD. Optical images of trench JBS diodes with fin-width of 0.6μm devices **(b)** before breakdown, and **(c)** post-breakdown.

The hard breakdown voltage of the trench JBSDs ranged from 1.8-2kV, while the planar SBDs and HJDs showed significantly lower breakdown voltages of 566V and 948V, respectively. This shows the effectiveness of the trench design and the SM-JTE compared to that of the planar devices without edge termination. The optical images of the devices pre-breakdown and post-breakdown are shown in Fig. 5 (b), (c), respectively. Most devices broke down due to high electric fields at the trench-corner or at the edge of the SM-JTE edge termination (discussed later). The trench width of the JBSDs showed no significant impact on the breakdown voltage as shown in Fig. 5 (a). This is likely due to the breakdown occurring at the trench corners where the peak field only shows a marginal increase with increasing fin width (Fig. 6 (b)) or at the edge of the SM-JTE, where the field is independent of the fin width. Furthermore, from the TEM images we observe PR residue at the trench corners (between the metal and NiO) which likely also compromised the breakdown field of the $NiO/Ga_2O_3$ junction at the trench corner. An optimized process would enable further enhancement of the overall junction breakdown field and the breakdown voltage of the JBSDs.

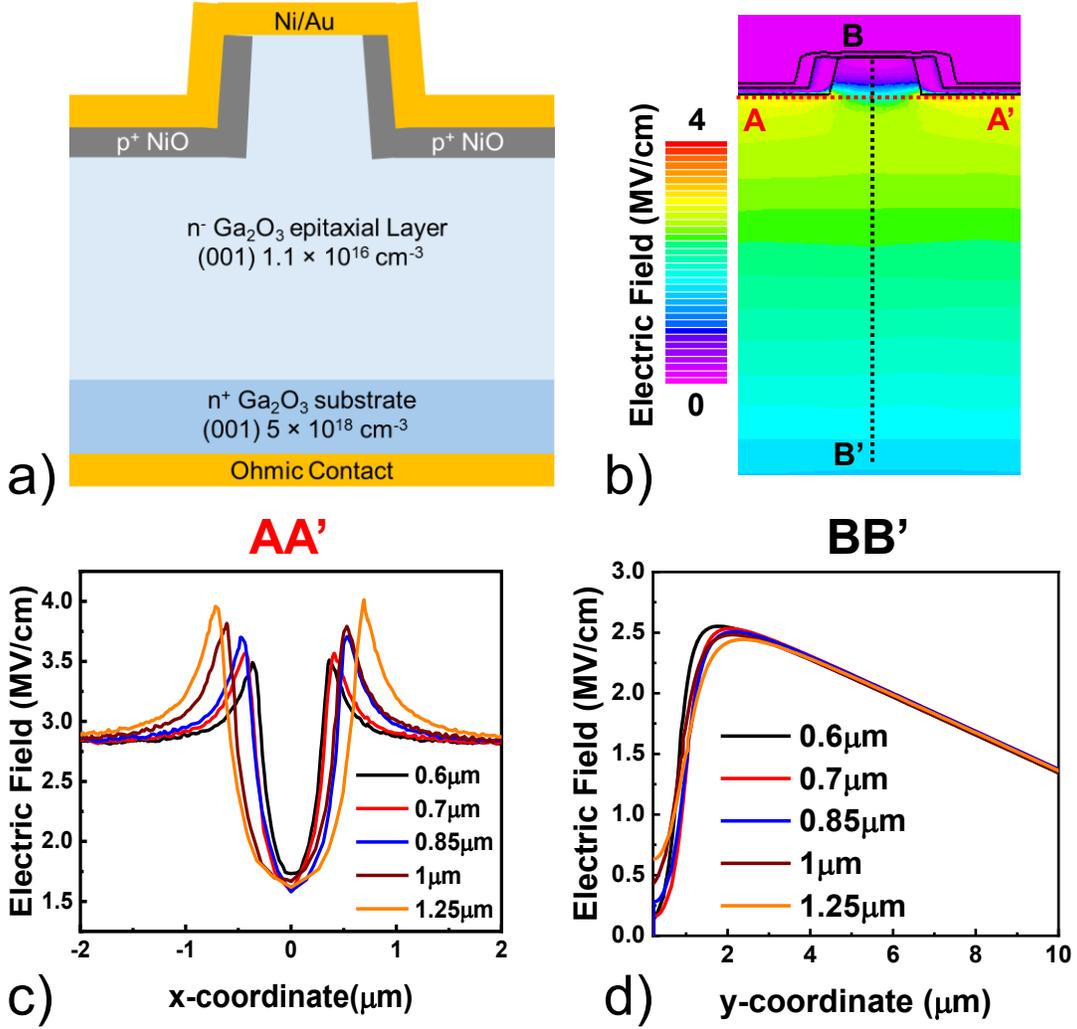

Fɪɢ. 6. TCAD silvaco simulation for trench JBSD with fin-width of 0.6μm at 2kV reverse bias is shown. **(a)** Schematic of a single-fin from trench JBS diodes. **(b)** Electric field contours showing 2 different cutlines AA' (horizontal line through Ga$_2$O$_3$ just at the NiO/Ga$_2$O$_3$ interface) and BB' (vertical line through Ga$_2$O$_3$ from top to bottom). **(c)** Plot of electric field vs x-coordinate (showing electric field distribution along AA'), and **(d)** Plot of electric field vs y-coordinate (showing electric field distribution along BB').

TCAD simulations for the trench diodes were performed using Silvaco to get additional insights into the field distribution within the device. The doping concentration in the NiO layer used is 4x10$^{18}$ cm$^{-3}$ which is estimated from C-V measurements of NiO/Ga$_2$O$_3$ diodes (fabricated on n$^+$ Ga$_2$O$_3$ substrate co-loaded during NiO sputtering). TCAD simulation results for trench JBSDs with varying fin-widths are shown in Fig. 5. The schematic of trench JBSD is shown in Fig. 6(a). The mesa angle at the fin bottom is about 83°. The electric field contour plot is shown in Fig. 6(b) and the electric field profile along cutlines AA' and BB' is shown in Fig. 6(c)-(d), respectively. At a reverse bias of 2 kV, the peak electric field at the trench corner varies between 3.5 and 4 MV-cm$^{-1}$ for fin-widths ($W_f$) between 0.6μm and 1.25μm, respectively. Due to the trench design, we are able to reduce the electric field at the metal/Ga$_2$O$_3$ interface to <0.6 MV-cm$^{-1}$ for all fin-widths used in this work (see Fig. 6(d)). This RESURF design prevents the premature breakdown due to high electric fields at the metal/ semiconductor junction.

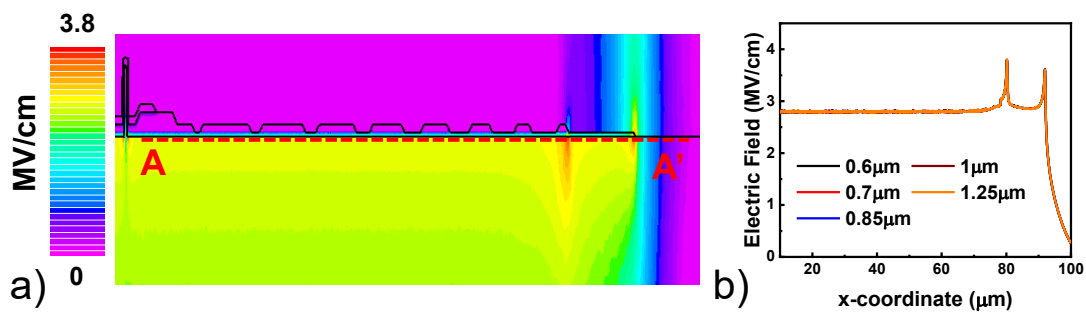

FIG. 7. TCAD silvaco simulations for trench JBSDs with SM-JTE edge termination at a reverse bias of 2kV. **(a)** Electric field contour at a reverse bias of 2 kV(horizontal line inside Ga$_2$O$_3$ at interface of NiO/Ga$_2$O$_3$), **(b)** Electric field along the cutline A-A' (see (b)) for devices with different fin-widths.

We also performed complete device TCAD simulations incorporating both the trench-structure and the SM-JTE edge termination as shown in Fig. 7. The electric field distribution along the horizontal cutline AA' is shown in Fig. 7(b) where we see 2 peak electric fields, one at the edge of JTE rings, and the other one at the edge of thin JTE layer, which is approximately 3.8 MV-cm$^{-1}$ (note that the peak electric field occurring at fin bottom is shown in Fig. 6).

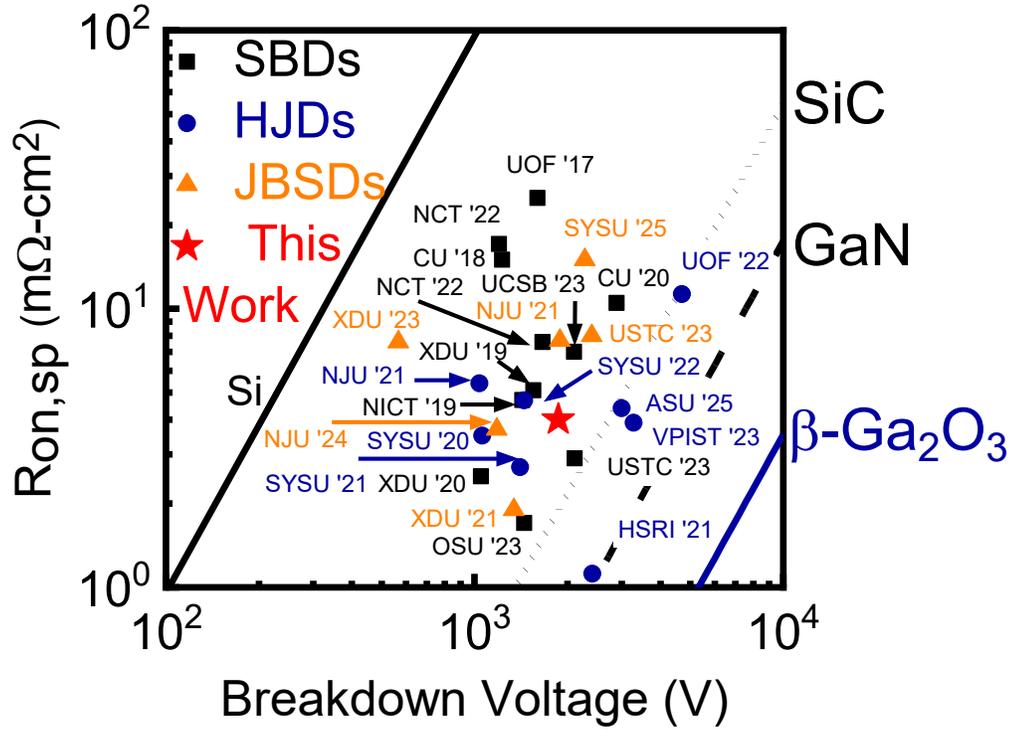

FIG. 8. Benchmarking of our best device with state-of-the-art Ga$_2$O$_3$ vertical SBDs, HJDs, and JBSDs.

The benchmark comparison of our trench JBSDs with prior reports of SBDs, HJDs, and JBSDs in terms of on-state resistance and breakdown voltage is shown in Fig. 8. The best performing JBSDs (W$_f$=0.85μm) shows a power figure of merit of 0.7 GW-cm$^{-2}$ (V$_B$~ 1867V and R$_{on,sp}$~4mΩ-cm$^2$) which is comparable to previous reports of vertical Ga$_2$O$_3$ diodes.[24,25,27–46]

## IV. CONCLUSION

In conclusion, we demonstrated trench junction barrier Schottky diodes integrated with Space Modulated-Junction termination extension achieving kV-class reverse breakdown performance and low turn on voltage close to 1 V. An effective tri-layer mask process with $Ni/SiN_x/SiO_2$ was used to etch the trenches and for deposition and lift-off of NiO from the top-side of the trenches. The reverse breakdown for the best performing diodes was around 2kV which resulted in unipolar PFOM of ~0.7 GW-cm$^{-2}$. These results exhibit the great potential of $NiO/Ga_2O_3$ junction barrier schottky diodes for low loss kilovolt class power electronics applications.

## ACKNOWLEDGMENTS

This work was supported by NSF ECCS-2336397 and in part by ULTRA, an EFRC center by DOE, Office of Science, Basic Energy Sciences under award No DE-SC0021230. The work in ASU NanoFab was supported in part by the National Science Foundation award ECCS-2025490. We acknowledge the use of facilities within the Eyring Materials Center at Arizona State University supported in part by NNCI-ECCS-1542160.